\documentclass[aps,prl,twocolumn,showpacs,groupedaddress]{revtex4}
\usepackage{epsfig}

\begin{document}
\title{Field-tuning of the electron and hole populations in the ruthenate 
Bi$_3$Ru$_3$O$_{11}$} 
\author{Wei-Li Lee,$^1$ M. K. Haas,$^2$ G. Lawes,$^3$ A. P. Ramirez,$^3$ R. J. 
Cava,$^{2,4}$ and N. P. Ong$^{1,4}$}

\affiliation{\textit{$^1$Department of Physics, $^2$ Department of Chemistry, Princeton 
University, Princeton, New Jersey 08544, U. S. A.\\ $^3$ Los Alamos National Laboratory, 
Los Alamos, New Mexico 87544, U.S.A.\\ $^4$Princeton Materials Institute, Princeton 
University, Princeton, New Jersey 08544, U. S. A.}}

\date{\today}
\begin{abstract}
Experiments on the Hall coefficient $R_H$ and heat capactity $C$ reveal an unusual, 
compensated electronic ground state in the ruthenate Bi$_3$Ru$_3$O$_{11}$.  At low 
temperature $T$, $R_H$ decreases linearly with magnetic field $|H|$ for fields larger 
than the field scale set by the Zeeman energy.  The results suggest that the electron 
and hole populations are tuned by $H$ in opposite directions via coupling of the spins 
to the field.  As $T$ is decreased below 5 K, the curve $C(T)/T$ vs. $T^2$ shows an 
anomalous flattening consistent with a rapidly growing Sommerfeld parameter $\gamma(T)$.  
We discuss shifts of the electron and hole chemical potentials by $H$ to interpret the 
observed behavior of $R_H$.
\end{abstract}
\maketitle
\section{Introduction}
The layered ruthenates have gained increased attention because of the discovery of 
superconductivity with triplet spin pairing in Sr$_2$RuO$_4$~\cite{Maeno}, 
metamagnetism~\cite{Perry} as well as field-tuned quantum critical 
behavior~\cite{Grigera} in Sr$_3$Ru$_2$O$_7$, and unusual ferromagnetism in 
SrRuO$_3$~\cite{Longo,Allen}.  Recently, a ruthenate from a different structural family 
La$_4$Ru$_6$O$_{19}$ has gained prominence because it exhibits non-Fermi liquid behavior 
at low temperatures~\cite{Khalifah}.  This oxide has the KSbO$_3$ structure which 
consists of a three dimensional network of edge-sharing and corner-sharing RuO$_6$ 
octahedra.  In this structure, a network of short Ru-Ru bonds co-exists with a network 
of Ru-O-Ru bonds.   The mix of nearly localized states centered on the short (2.49 $\rm 
\AA$) Ru-Ru bonds and delocalized states derived from the Ru-O-Ru bonds is rare in 
transition-metal oxides.  The interaction between them is analogous to that between $f$ 
electrons and $s$ electrons in heavy-fermion materials.  Distinct signatures of 
anomalous behavior are observed in La$_4$Ru$_6$O$_{19}$.  The heat capacity $C(T)$ 
displays a $T\log T$ profile below 1 K in zero magnetic field.  The resistivity $\rho$ 
shows a $T$-linear dependence below 30 K that extrapolates to zero as $T\rightarrow 0$ 
(with no measurable residual resistivity), which is a clear signature of non-Fermi 
liquid behavior.  By contrast, these anomalous properties are absent in the closely 
related compound La$_3$Ru$_3$O$_{11}$ in which the Ru-Ru distance is much longer (2.99 
$\rm \AA$).  The anomalous electronic properties of La$_4$Ru$_6$O$_{19}$ are crucially 
dependent on the presence of the narrow peak in the density-of-states derived from the 
Ru-Ru dimers.  

We have synthesized a third member~\cite{He} of this family Bi$_3$Ru$_3$O$_{11}$ in 
which the Ru-Ru distance (2.61 $\rm \AA$) lies between those of the previous two.  Heat 
capacity measurements reveal that, at low $T$, the Sommerfeld parameter $\gamma$ 
increases as $T$ decreases below 5 K.  The Hall coefficient $R_H(H)$ reveals that the 
ground state is compensated but the relative electron and hole populations are highly 
sensitive to the field $\bf H$.   In relatively weak $H$, $R_H(H)$ decreases linearly 
with increasing $|H|$, implying that the electron and holes populations are tuned in 
opposite directions.

\section{Heat Capacity and resistivity}
Polycrystalline samples of Bi$_3$Ru$_3$O$_{11}$ were prepared by pressing high-purity 
Bi$_3$Ru$_3$O$_{11}$ powder under 50 kbar at 700 C for 2 h, and then annealing overnight 
at 900 C under ambient pressure.  Details of the sample preparation, structural and 
thermodynamic measurements are reported elsewhere~\cite {Haas}.  
\begin{figure}[h]
\includegraphics[width=8cm]{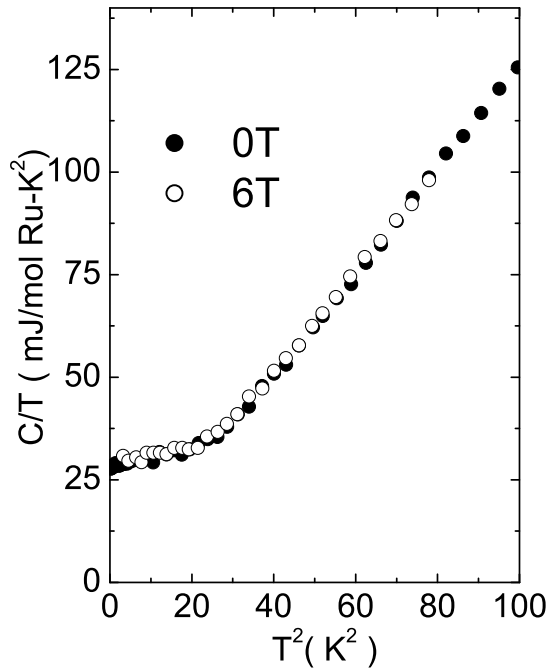}
\caption{\label{heatcap} Plot of the low-$T$ heat capacity $C(T)$ as $C(T)/T$ vs. $T^2$ 
in Bi$_3$Ru$_3$O$_{11}$ in zero field (solid circles) and in a 6-T field (open).  Below 
5 K, the curves flatten out to yield a relatively large Sommerfeld constant $\gamma =$ 
28 mJ/mol Ru-K$^{2}$.}
\end{figure}
\begin{figure}[h]
\includegraphics[width=8cm]{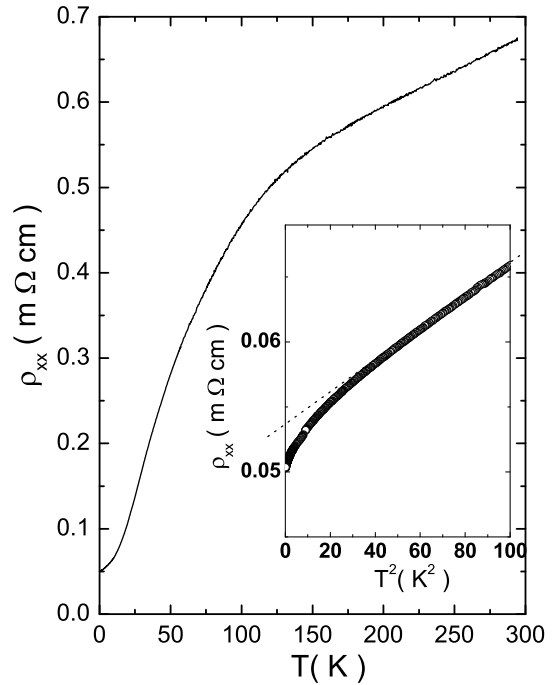}
\caption{\label{rho}The $T$ dependence of the resistivity $\rho$ in 
Bi$_3$Ru$_3$O$_{11}$.  Between 5 and 25 K, $\rho$ fits well to a $T^2$ dependence, but 
shows a slight downwards deviation below 5 K (inset).
}
\end{figure}
The low-temperature heat capacity $C(T)$ of Bi$_3$Ru$_3$O$_{11}$ is plotted as $C(T)/T$ 
vs. $T^2$ in Fig. \ref{heatcap}.  The curve starts out nominally linear above 5 K with 
an apparent extrapolated intercept at the value `$\gamma$' = 5.3 mJ/mol Ru K$^2$.  
However, below 5 K, the curve flattens out to a much less $T$-dependent line that 
extrapolates at $T=0$ to the value $\gamma$ = 28 mJ/mol Ru K$^2$, which is $\sim$3 times 
smaller than in La$_4$Ru$_6$O$_{19}$.  The value of $\gamma$ is enhanced by a factor of 
4-5 over that in conventional metals, but less than that in heavy-fermion systems (for 
which $\gamma$ $\sim$ 300-1500 mJ/mol K$^2$).  The highly unusual feature of $C(T)/T$ is 
the weak $T$ dependence of the curve at low $T$.  The straight line corresponding to the 
usual phonon contribution varying as $T^3$ is absent.  We interpret the relative flat 
curve below 5 K as a gradual crossover to an unusual electronic state in which the 
carrier effective mass $m^*$ rises by a factor of $\sim 5$ to raise $\gamma$ to its 
value at $T=0$.  The unusual nature of the ground state becomes apparent in the Hall 
results reported in the next section.

As in La$_4$Ru$_6$O$_{19}$ and La$_3$Ru$_3$O$_{11}$, the resistivity $\rho$ is metallic.  
In our polycrystalline sample, however, the residual resistivity ratio (RRR $\sim 13$) 
is smaller because of significant scattering from disorder and impurities.  As $T$ falls 
from 300 to 120 K, $\rho$ initially decreases linearly with $T$ (Fig. \ref{rho}).  Below 
100 K, $\rho$ drops rapidly to the residual value $\rho_0$.   The variation below 25 K 
is nearly linear in $T^2$ but deviates below 5 K (see inset to Fig. \ref{rho}).  
\begin{figure}[h]
\includegraphics[width=8cm]{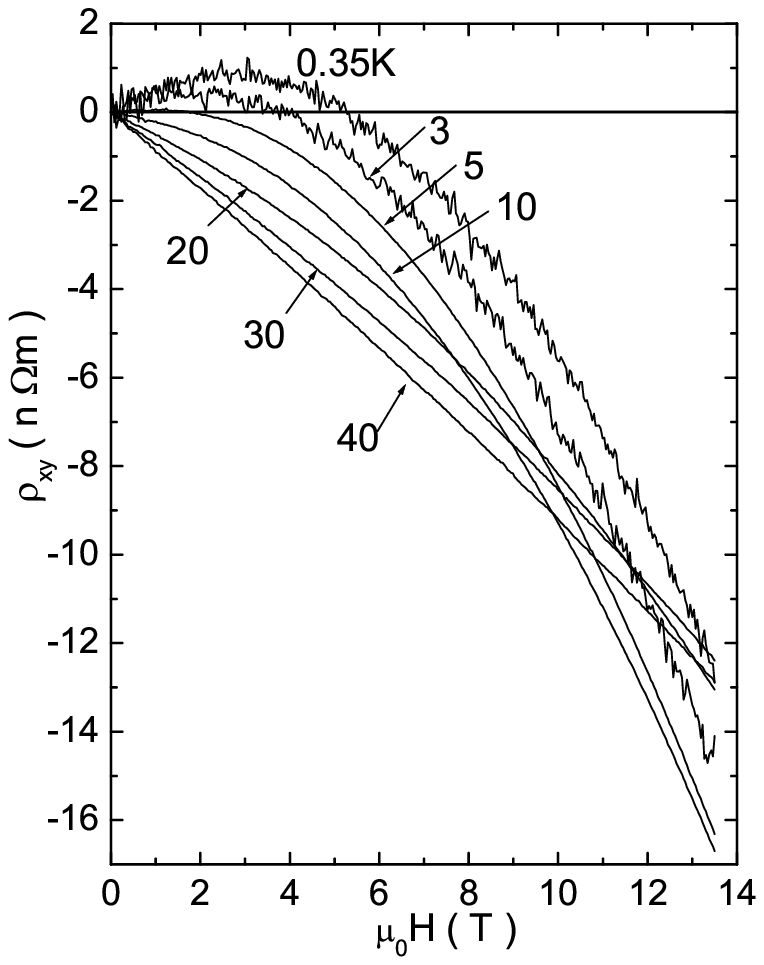}
\caption {\label{rhoxy} Curves of $\rho_{xy}$ vs. $H$ in Bi$_3$Ru$_3$O$_{11}$ showing 
unusually pronounced curvature vs. $H$ below 50 K.  The weak-field slope $\partial 
\rho_{xy}/\partial H$ increases steeply with $T$ as $T\rightarrow 0$.  Above 50 K, 
$\rho_{xy}$ is strictly linear in $H$ to 14 T.
} 
\end{figure}
\begin{figure}[h]
\includegraphics[width=8cm]{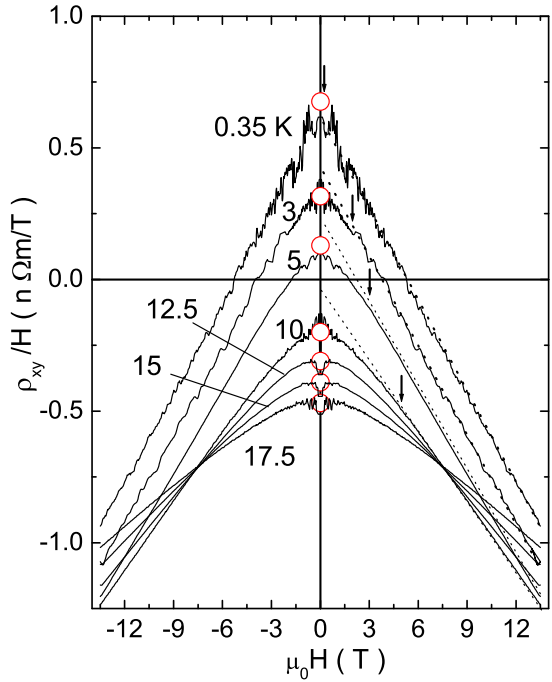}
\caption{\label{RHvsH} Replot of $\rho_{xy}$ as $R_H(H) = \rho_{xy}/H$ vs. $H$.  At each 
$T$, we have reflected the \emph{same} curve about $H = 0$ to emphasize that $R_H$ is 
symmetric in $H$.  At large $H$, $R_H$ is strictly linear in $H$ (dashed lines), but 
below the Zeeman field scale $H_Z$ (arrows), $R_H(H)$ deviates from the dashed line 
because of thermal broadening.  Open circles represent $R^0_H$.
}
\end{figure}
The bulk susceptibility $\chi$ measured in a SQUID magnetometer with $H$ = 1 T rises 
gradually to the value $3 \times 10^{-4}$\ emu/G.mol Ru as $T$ is cooled from 400 K to 
25 K, consistent with Pauli susceptibility~\cite{Haas}.  Below 25 K, however, we observe 
a steep increase suggestive of spin fluctuations.  However, there is little evidence for 
local moments or magnetic ordering in $\chi$ (at 5 K, $\chi$ remains independent of $H$ 
up to 5 T and corresponds to 0.004 $\mu_B$ per Ru ion at 5 T).  
\section{Hall effect and Magnetoresistance}
The Hall effect and magnetoresistance together provide a powerful way to probe the 
nature of the electronic state  at low $T$.  At temperatures 50 to 300 K, we find that 
the Hall resistivity $\rho_{xy}$ is strictly linear in $H$ (data not shown).  Above 200 
K, the Hall coefficient $R_H = \rho_{xy}/H$ is positive and approaches the 
$T$-independent value $0.72 \times 10^{-9}$ m$^3$/C (corresponding to a Hall number $n_H 
= 8.7\times 10^{21}$ cm$^{-3}$).  At 110 K, $R_H$ changes sign, implying that 
electron-like and hole-like Fermi Surfaces (FS) are present.  Interestingly, below 50 K, 
the curve of $\rho_{xy}$ vs. $H$ rapidly acquires pronounced curvature in fields of 1-3 
T (Fig. \ref{rhoxy}).  Such highly pronounced curvature, strikingly unusual in a system 
with such a short mean-free-path $\ell$ (500-900 \AA), signals that the applied field 
strongly affects the electronic state itself.  Below 50 K, the observed Hall signal is 
the combined effect of $H$ exerting a Lorentz force on the carriers and simultaneously 
altering the ground state.

To verify this, we divide $\rho_{xy}(H)$ by $H$ to define the \emph{field-dependent} 
Hall coefficient $R_H(H) = \rho_{xy}/H$ (Fig. \ref{RHvsH}).  This removes the leading 
$H$-linear factor in $\rho_{xy}$ leaving an $R_H(H)$ that is even in $H$ (we have folded 
the curves about the axis $H = 0$ in Fig. \ref{RHvsH} to emphasize this point).  As 
indicated by the dashed lines, $R_H(H)$ is linear in $|H|$ at high fields, but as 
$H\rightarrow 0$, it deviates downwards consistent with a `rounding' in weak fields (we 
associate this rounding with thermal broadening).  As discussed below, these features 
strongly suggest that the field increases the electron-like population while diminishing 
the hole-like population via Zeeman coupling to the spins of the carriers.  

At each $T$, the curves in Fig. \ref{RHvsH} are characterized by two parameters, the 
zero-field Hall coefficient $R^0_H(T)$ and the slope of the dashed lines ${\cal P}(T) 
\equiv |dR_H/dH|$ (evaluated at 14 T).  At the lowest $T$, $R^0_H$ falls steeply with 
increasing $T$ (solid symbols in Fig. \ref{RH0}).  As the carrier lifetimes are not $T$ 
dependent in the impurity scattering regime, this rapid variation cannot arise from 
changes to $\ell$.  Instead, it comes from changes to the density of states (DOS) in the 
electron and hole bands at finite $T$.  The second parameter ${\cal P}(T)$ measures the 
rate of decrease in $R_H$ with increasing $H$.  As $T$ is decreased from 50 K, ${\cal 
P}(T)$ grows gradually  (open symbols in Fig. \ref{RH0}), but saturates to a constant 
near 5 K.  The saturation of ${\cal P}(T)$ at low $T$ recalls the behavior of $C(T)/T$ 
in Fig. \ref{heatcap}.  Before discussing the Hall results further, we consider the 
magnetoresistance (MR).

Figure \ref{MR} shows the transverse MR $[\Delta\rho/\rho]_{\perp}$ (measured with $\bf 
H\perp I$ with $\bf I$ the applied current) and the longitudinal MR 
$[\Delta\rho/\rho]_{||}$ (measured with $\bf H|| I$) at low $T$.   As $\bf H|| I$ in the 
longitudinal MR geometry, the field couples onto to the spin degrees of freedom (either 
of the carrier or local moments that scatter the carriers).  At each $T$, the orbital 
component may be isolated by subtracting the longitudinal MR signal from the 
transverse~\cite{Harris}.  In plotting $[\Delta\rho/\rho(0)]_{orb}$ against $H/\rho(0)$ 
(Kohler plot), we find that curves taken at different $T$ all collapse together.  The 
Kohler scaling confirms that the orbital component of the MR arises entirely from the 
effect of the Lorentz force on the electron trajectory (classical MR).  The magnitude of 
$[\Delta\rho/\rho(0)]_{orb}$ places an upper bound for the value of $\ell_0$ of 900 \AA, 
which is rather short, and consistent with the modest RRR.

\section{Field-effect on carrier populations}
\begin{figure}[h]
\includegraphics[width=8cm]{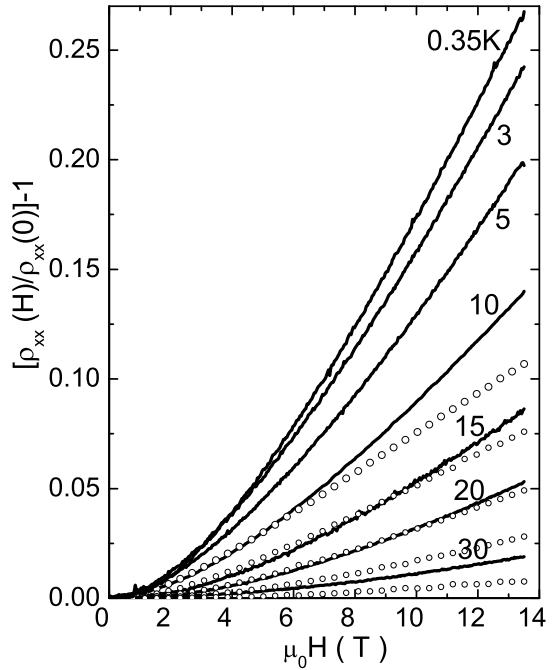}
\caption {\label{MR} The transverse magnetoresistance $[\Delta\rho/\rho]_{\perp}$ ($\bf 
H\perp I$) at selected $T$ (solid curves).  Curves of the longitudinal MR 
$[\Delta\rho/\rho]_{||}$ ($\bf H|| I$) at $T$ at 5, 10, 15, 20 and 30 K, are displayed 
as small open circles.}
\end{figure}

The Hall results in Fig. \ref{RHvsH} imply that an applied field alters the 
\emph{relative} electron and hole populations linearly at large $H$.  The MR results 
(Fig. \ref{MR}), however, show that the change to the combined carrier population is 
negligible even at our lowest $T$.  Since the decrease in $R_H(H)$ is rigorously linear 
in $H$ at large enough fields, the changes induced by field must arise from the coupling 
of $H$ to the carrier spins by their Zeeman energy.  For the field to be effective, the 
Zeeman energy must clearly exceed $k_BT$, i.e. $H$ must exceed the Zeeman field scale 
$H_Z = k_BT/g\mu_B$ where $k_B$ is Boltzmann's constant, $g$ is the Lande g-factor and 
$\mu_B$ the Bohr magneton.  Examination of the curves of $R_H$ in weak fields shows that 
this is indeed the case.  At each $T$, $R_H$ deviates from the dashed lines when $H<H_Z$ 
(arrows in Fig. \ref{RHvsH}).
\begin{figure}[h]
\includegraphics[width=8cm]{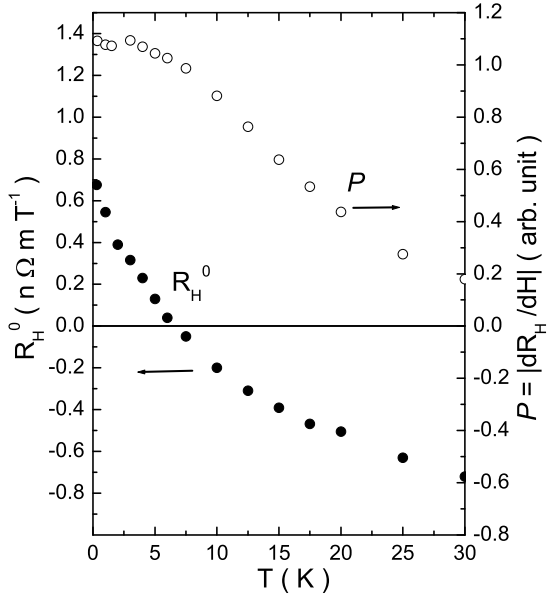}
\caption{\label{RH0} The $T$ dependence of the zero-field Hall coefficient $R^0_H(T) = 
\rho_{xy}/H$ ($H\rightarrow 0$) (solid symbols) and ${\cal P}(T) \equiv |dR_H/dH|$ 
evaluated at 14 T (open symbols) below 30 K.  At low $T$ (0.35-5 K), the steep decrease 
of $R^0_H$ arises from changes to the DOS.  The parameter ${\cal P}$ is the change in 
$R_H(H)$ induced by unit $H$. }
\end{figure}
For simplicity, let us assume a two-band model, in which $\sigma^i$ and $\sigma^i_H$ 
are, respectively, the conductivity and Hall conductivity in the electron-like band ($i 
= e$) or hole-like band ($i = h$).  In the impurity scattering regime, the 
mean-free-path is a $\bf k$-independent length $\ell_0$ that has the same value in both 
bands.  We may then write for the $i^{th}$ band~\cite{Ziman}
\begin{equation}
\sigma^i = g{\cal S}_i\ell_0, \quad\quad\quad
\sigma^i_{H} = g{\cal S}_i \frac{eH}{\hbar k_{F,i}}\ell_0^2 \equiv \pm \ g{\cal R}_i 
H\ell_0^2,
\label{cond}
\end{equation}
where $g = e^2(12\pi^3\hbar)^{-1}$, ${\cal S}^i$ is the `transport weighted' Fermi 
Surface area of the $i^{th}$ band, $k_{F,i}$ the average Fermi wavevector, and the +(-) 
sign applies to $\sigma^h_H$ ($\sigma^e_H$).  For isotropic bands, the quantity ${\cal 
R}_i$ may be regarded as proportional to DOS of the $i^{th}$ band.  In this limit, $R_H$ 
is independent of $\ell_0$, viz. 
\begin{equation}  
R_H(H) = \frac{(\sigma^e_{H} + \sigma^h_{H})}{[H(\sigma^e + \sigma^h)^2]} \rightarrow 
\frac{{\cal R}_h(H) - {\cal R}_e(H)}{ g({\cal S}_h + {\cal S}_e)^2},
\label{RHH2}
\end{equation}

As the orbital MR is weak, we may ignore the effect of $H$ on the denominator in Eq. 
\ref{RHH2}.  The linear decrease in $R_H$ at large $H$ then implies that both ${\cal 
R}_e$ and ${\cal R}_h$ must vary linearly with $H$, but with opposite signs.  At $T = 
0$, we have ${\cal R}_i(H) = {\cal R}_i(0)[1 - \alpha_i |H|]$, with the parameters 
$\alpha_e<0$ and $\alpha_h >0$.  In the Drude approximation, ${\cal R}_i$ is 
proportional to the carrier density in the $i$ band.  As soon as $H$ is non-zero, the 
electron population increases linearly while the hole population decreases.  

At finite $T$, thermal broadening makes this field effect insignificant until $H$ 
exceeds $H_Z$, as noted.  In high fields, the rate at which $H$ alters the relative 
populations, measured by ${\cal P}$, is independent of $T$ below 5 K (Fig. \ref{RH0}).  
Above 5 K, the field sensitivity diminshes gradually, becoming undetectable at 50 K.

\section{Discussion}
In analyzing the $\rho_{xy}$ curves, we have ignored the possibility of skew scattering 
contributions to the Hall current from scattering off local moments or spin 
fluctuations.  In heavy fermion systems, the spin fluctuations of local moments 
typically dominate the Hall effect, and curvature in $\rho_{xy}$ vs. $H$ may be observed 
if magnetic ordering is present.  For e.g. in CeAl$_2$, a metamagnetic transition near 5 
T strongly affects the curve of $\rho_{xy}$ vs. $H$~\cite{Haen}.  However, as discussed 
above, the evidence from $\chi$ for local moments in Bi$_3$Ru$_3$O$_{11}$ is quite weak.  
The estimated moment on each Ru ($<0.004 \mu_B$ at 5 K) is far too feeble to produce 
observable magnetic scattering.  More importantly, the behavior of $\rho_{xy}$ with $H$ 
(and $T$) in skew scattering has a well-studied characteristic form arising from domain 
rotation.  The linear dependence of $\rho_{xy}/H$ in Fig. \ref{RHvsH} is incompatible 
with this form.  The available evidence persuades us that the anomalous behavior in Fig. 
\ref{RHvsH} is unrelated to Hall currents from skew scattering off magnetic moments.

In a paramagnetic one-band metal, the shift in field of the chemical potential $\mu_{+}$ 
($\mu_{-}$) for spin-up (spin-down) electrons changes the spin sub-populations by 
$\delta n_{\pm} = \pm\frac12 {\cal N}_F\mu_BH$, to give the familiar Pauli magnetization 
$M = {\cal N}_F \mu_B^2H$ (here ${\cal N}_F$ is the total DOS at the Fermi level).  
However, $R_H$ is unaffected because the Hall effect is sensitive to the total 
population, not $\delta n_{\pm}$.  For a system with holes and electrons, the argument 
applies independently to each FS, so $R_H$ remains unchanged.  The data in Fig. 
\ref{RHvsH} require an unusual ground state in which the chemical potentials of the 
holes and electrons ($\mu_h$ and $\mu_e$, respectively) behave as if they are 
Zeeman-coupled to the field with opposite signs, and shift just like $\mu_{\pm}$, viz.
\begin{equation}
\mu_h(H) = \mu_h(0) + \frac12 g\mu_BH, \quad\quad \mu_e(H) = \mu_e(0) - \frac12 g\mu_BH. 
\label{mu}
\end{equation}
The implication is that, in a field, the hole and electron bands are each 
spin-polarized, but in opposite directions.  In a weak $H$ at $T = 0$, $\mu_e$ shifts 
downwards (this is dictated by the observed decrease in $R_H(H)$) leading to an increase 
in the electron population, while $\mu_h$ shifts upwards leading to a decreased hole 
population.  As $\rho$ is virtually unchanged, the net population change is nearly zero.  

The inferred coupling between the hole and electron bands is broadly reminiscent of 
excitonic condensates~\cite{Rice} involving the pairing of holes and electrons in a 
compensated semi-metal.  Recent work~\cite{Varma} has explored various magnetic ground 
states.  However, our observations do not seem to have been predicted.  Moreover, we 
caution that the opening of an energy gap has not been observed.  The nearest feature to 
an order parameter seems to be ${\cal P}$, which appears near 50 K and gradually 
increases to saturate at 5 K, a crossover behavior reflected in the heat capacity.  
Hopefully, the results reported will motivate a search for a correlated state that 
reproduces the observations in Fig. \ref{RHvsH}.

\acknowledgments{The research reported is supported by a MRSEC grant from the U.S. 
National Science Foundation (DMR 0213706). We thank M. Sato for drawing our attention to 
Ref. \cite{Fujita}.}

\end{document}